\documentclass[twocolumn]{aastex63}
\usepackage{amssymb,graphicx,verbatim}
\usepackage{natbib}
\newcommand{\rev}[1]{{\color{red!50!black} #1}}
\setcitestyle{notesep={ }} 
\newcommand{\arcs}{\ensuremath{^{\prime\prime}}}
\newcommand{\arcm}{\ensuremath{^{\prime}}}
\newcommand{\sDM}{\ensuremath{\frac{\sigma_{\rm DM}}{m}}}
\renewcommand{\rev}[1]{#1}

\begin{document}

\title{Line of Sight Bias in Dark Matter Inferences from Galaxy Cluster Mergers}
\shorttitle{Line of Sight Bias in Galaxy Cluster Mergers}

\author[0000-0002-0813-5888]{David Wittman}
\affiliation{Department of Physics and Astronomy, University of California, Davis, CA 
  95616 USA}
\author[0009-0004-6902-4649]{Scott Adler}
\affiliation{Department of Physics and Astronomy, University of California, Davis, CA 
  95616 USA}
\author[0000-0002-6217-4861]{Rodrigo Stancioli}
\affiliation{Department of Physics and Astronomy, University of California, Davis, CA 
  95616 USA}


\begin{abstract} In collisions of galaxy clusters, the lack of
  displacement between dark matter and galaxies suggests that the dark
  matter scattering depth is small. This yields an upper limit on the
  dark matter cross section if the dark matter column density is
  known.  We investigate a bias in such constraints: the measured
  column density (along the line of sight, using gravitational
  lensing) is lower than that experienced by a dark matter particle,
  as follows.  Dark matter halos are triaxial and generally collide
  along their major axes, yielding a high scattering column
  density---but the merger is obvious only to observers whose line of
  sight is nearly perpendicular to that axis, yielding a low observed
  column density. We trace lines of sight through merging halos from
  the BigMDPL n-body simulation, both with and without mock
  observational effects. We find that a hypothetical skewer through
  the center of the halo along the merger axis (more precisely, along
  the current separation vector of the two halos) has \rev{nearly}
  twice the column density of a typical line of sight.  With weak
  lensing measurements, which involve some spatial averaging, this
  ratio is reduced to 1.25, suggesting that existing constraints on
  the scattering cross section are biased high by about 25\%.
\end{abstract}

\section{Introduction}\label{sec-intro}

Dark matter (DM) is the dominant form of matter in the universe, but little
is known about its particle properties. One key question is to what
extent, if any, DM particles can scatter off each other. The
astrophysical effects of such scattering, also known as
self-interacting dark matter (SIDM), were first discussed by
\cite{Spergel00}, who also presented rough upper limits on the
scattering cross section per unit mass \sDM\ based on those effects.  In many
particle models, \sDM\ will be velocity-dependent, so constraints from
different astrophysical environments should be considered in the
context of the typical particle velocity in that environment
\citep{Kaplinghat16}. The highest velocities probed are in the mergers
of galaxy clusters, with relative speeds of $\sim3000$ km/s
\citep{Markevitch04,Randall2008,Harvey15,Robertson17Bullet,Mismeasure2018}.

\citet{Markevitch04} identified the Bullet cluster as a
post-pericenter snapshot of a nearly head-on collision between two
galaxy clusters.  A hallmark of recent pericenter passage is a
substantial separation between gas and galaxies (later called
\textit{dissociation} by \citet{Dawson11}) due to momentum
exchange between the gas distributions around the time of pericenter.
\citep{Markevitch04} argued that the \textit{lack} of separation
between DM and galaxies implied that the DM scattering depth
$\tau = \sDM\Sigma < 1$, where $\Sigma$ is the surface mass density
encountered by a DM particle of one halo as it passes through the
other halo. Hence $\Sigma^{-1}$ yields an upper limit on \sDM\ if no
significant galaxy-DM separation is observed. \citet{Harvey14}
generalized this argument to yield an estimate of \sDM\ if a
separation \textit{is} observed.

This paper explores a bias in this method. In practice $\Sigma$ is
measured along the line of sight (LOS), but for purposes of the DM
constraint it should be measured along the merger trajectory, or along
the current separation vector (CSV) as a proxy. Because halos are
triaxial \citep{Jing2002} and preferentially aligned with neighboring
halos \citep{Binggeli1982,Plionis1994,Kasun_2005,Smargon2012}, the
surface mass density probed along the CSV, $\Sigma_{\rm CSV}$, will
tend to be larger than the surface mass density probed along a random
LOS, $\Sigma_{\rm los}$. Furthermore, the merging systems used for
these constraints tend to be those in which the LOS is perpendicular
to the CSV, such that physical separations would be visible as angular
separations.  This implies that $\Sigma_{\rm los}$ is measured along a
triaxial halo's intermediate or minor axis, biasing it low compared to
a random LOS. If the ratio
$\frac{\Sigma_{\rm los}}{\Sigma_{\rm CSV}}<1$, then the DM scattering
depth is underestimated and \sDM\ is overestimated.  This paper
examines dark matter-only n-body simulations to quantify the factor by
which \sDM\ is overestimated.

In \S\ref{sec-methods} we describe our methods, in \S\ref{sec-results}
we present the results, and in \S\ref{sec-discussion} we discuss the
impact of the results.  

\section{Methods}\label{sec-methods}

We approximate each halo as a triaxial ellipsoid, meaning that its
density $\rho$ is a function only of the scaled radial coordinate
\rev{$r\equiv\sqrt{(\frac{x}{a})^2 + (\frac{y}{b})^2 +
    (\frac{z}{c})^2}$} where $a$, $b$, and $c$ denote the semiaxis
lengths ($a>b>c$).  We assume for notational simplicity here a
Cartesian coordinate system aligned with the halo principal axes; in
practice, this can be achieved by a rotation from the coordinate
system of the simulation in which the halo is embedded.

\rev{\subsection{Central Skewers}}

The surface mass density $\Sigma$ along any skewer
piercing the \rev{center of the} ellipsoid can be expressed as the
skewer distance across the ellipsoid times an integral
$\int_0^\infty \rho(r)dr$ that depends \textit{only} on the density
profile. Therefore the ratio
$\frac{\Sigma_{\rm los}}{\Sigma_{\rm CSV}}$ cancels the latter factor
and involves only the ratio of distances across the ellipsoid. \rev{In
  this subsection we argue that this simple ratio is a useful
  thinking tool because it provides a bound on the size of the effect
  when integrated over larger surface areas, independent of the
  density profile or observational technique. Later in the paper, we
  perform more realistic estimates with mock weak gravitational
  lensing fits, which inherently average over a patch of sky.}

\begin{figure}
\centerline{\includegraphics[width=\columnwidth]{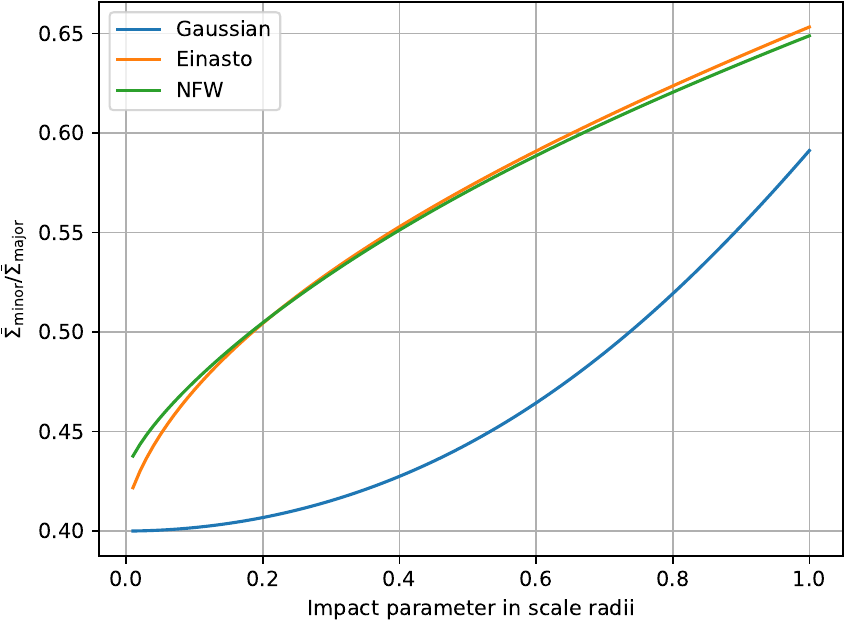}}
\caption{\rev{Mean surface mass density along the minor axis,
    $\bar{\Sigma}_{\rm minor}$, as a fraction of that along the major
  axis, $\bar{\Sigma}_{\rm major}$, as a function of projected radius
of the area enclosed, for a halo with minor axis ratio
$\frac{c}{a}=0.4$. The central skewer prediction of 0.4 provides a
lower bound independent of density profile.}}
\label{fig-sigmaskewer}
\end{figure}

\rev{Consider a LOS parallel to the major ($x$) axis, but displaced in
  $z$: Taylor expanding the density about $z=0$ produces a term
  proportional to $\frac{\rho^\prime(r)}{c^{2}}$. In contrast,
  expanding a minor-axis LOS about $x=0$ produces a term proportional
  to $\frac{\rho^\prime(r)}{a^{2}}$. Therefore, the density drops more
  quickly for LOS displaced from the major axis; including these
  non-central effects over a finite patch of sky will decrease
  $\bar{\Sigma}_{\rm major}$ faster than it decreases
  $\bar{\Sigma}_{\rm minor}$, where the bars indicate area
  averaging. Hence, skewers through the center provide a useful lower
  bound on the ratio of \textit{mean} surface mass densities averaged
  over finite areas. We confirm this by using the formalism of
  \citet{Heyrovsky24} to find
  $\frac{\bar{\Sigma}_{\rm minor}}{\bar{\Sigma}_{\rm major}}$ averaged
  over circles centered on the principal axes, for a variety of
  density profiles and circle sizes. Figure~\ref{fig-sigmaskewer}
  shows the result for a halo with $\frac{b}{a}=0.6$ and
  $\frac{c}{a}=0.4$, for three profiles: NFW \citep{NFW97}, Einasto
  \citep{Einasto1965}, and Gaussian. The scale radius of the Einasto
  profile was chosen to match the NFW.  The Gaussian profile is not
  physically motivated, but illustrates the behavior of an entirely
  different profile. With $\frac{c}{a}=0.4$, the central skewer
  argument predicts
  $\frac{\Sigma_{\rm minor}}{\Sigma_{\rm major}}=0.4$. Near the
  center, the Gaussian density profile is close to flat so the ratio
  remains close to 0.4 even when averaged over a substantial area. The
  other profiles are not flat near the center so
  $\frac{\bar{\Sigma}_{\rm minor}}{\bar{\Sigma}_{\rm major}}$
  increases more rapidly with averaging radius. In all cases the
  central skewer ratio of 0.4 provides a lower bound, and
  overestimates the departure from unity by less than a factor of
  two.}

\rev{Having established the utility of the central skewers, we now
  proceed to geometric definitions that will be used for both the
  skewer estimates and the mock weak lensing. At the end of this
  section we transition to describing the weak lensing simulations.}
  
\subsection{Geometric definitions}

\begin{figure}
\centerline{\includegraphics[width=0.9\columnwidth]{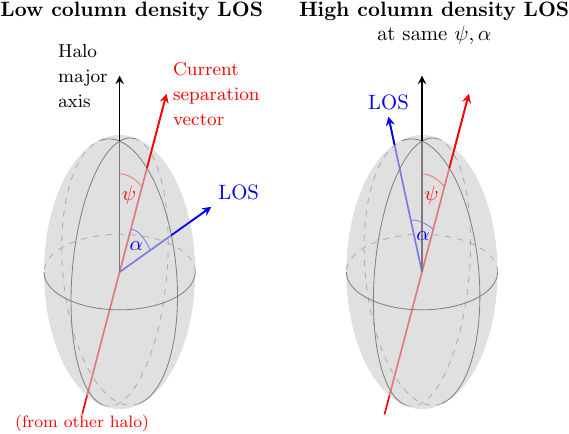}}
\caption{\rev{Geometric view.  $\psi$ separates the halo major axis
  and the current separation vector to the other halo in the merger (not shown
  here). Lines of sight (LOS) fall along a cone defined by the
  separation vector and $\alpha$. At fixed $\psi$ and $\alpha$, LOS
  can vary substantially in column density.}}
\label{fig-ellipsoid}
\end{figure}

Figure~\ref{fig-ellipsoid} illustrates the geometry used in this
paper. The black arrow defines the first principal axis (major
axis) of the triaxial halo, while the red arrow shows an
illustrative CSV (the other halo involved in the merger is not
shown). Halos tend to be aligned with their neighbors, so the angle
$\psi$ is generally small as shown. The LOS is drawn at a larger angle
$\alpha$ from the separation vector, because lines of sight near the
separation vector would not identify a merger as dissociative. There
are techniques for constraining $\alpha$ \citep[e.g.,][]{Analogs2018}
but these do not constrain the relationship between the LOS and the
halo principal axes. Hence, the LOS could be anywhere along a cone
with opening angle $\alpha$. Figure~\ref{fig-ellipsoid} shows two
possible LOS along this cone, illustrating that one involves a
considerably longer path through the halo hence a larger
$\Sigma_{\rm los}$.

We calculate the distance through the halo along the CSV, and along
all possible LOS (at a given $\alpha$) to obtain a distribution for
$\frac{\Sigma_{\rm los}}{\Sigma_{\rm CSV}}$ given a specific halo,
merger, and $\alpha$.  We can then marginalize over many halos and
mergers to obtain a distribution for
$\frac{\Sigma_{\rm los}}{\Sigma_{\rm CSV}}$ given $\alpha$.  The
selection of mergers could be tailored to match a specific observed
system as in \citet{Wittman19analogs}, but this paper derives results
for mergers in general.

\subsection{Simulation inputs and calculations} \label{sec-m-simulation}

We used the publicly available Big Multidark Planck
\citep[BigMDPL,][]{BigMDPL2016} simulation hosted on the
CosmoSim\footnote{\url{https://www.cosmosim.org/metadata/bigmdpl/}}
  website \citep{CosmoSim2013}. This dark matter simulation has a very
  large box size, $(2.5\ \textrm{Gpc}/h)^3$, which maximizes the
  number of merging halo pairs.  The mass of each particle in the
  simulation is $2.359 \times 10^{10} M_{\odot}$, so even the least
  massive clusters considered here ($M > 6 \times 10^{13}\,M_{\odot}$)
  have over 250 particles. The $\Lambda$CDM cosmological parameters
  were set to $h=0.6777$, $\Omega_\Lambda = 0.692885$, and
  $\Omega_m=0.307115$.

We use the catalog of halo pairs assembled in \cite{Analogs2018} and
\citet{Wittman19analogs}. Briefly, this is derived from the BigMDPL
Rockstar \citep{Rockstar2013} halo catalog by finding halo pairs that
have passed exactly one pericenter (within 300 kpc). This is in
principle a different catalog at each snapshot, but we focus on one
illustrative snapshot (number 74, redshift $0.1058$). We discarded halo pairs
with current separations $<0.25$ Mpc, as these were too overlapping
for Rockstar to derive robust shape parameters. Observed mergers with
current separations $<0.25$ Mpc are also rare because they are
observationally difficult to disentangle.

We use the \texttt{axis1\_[xyz]} attributes (square brackets denote
use of each enclosed letter) to determine the orientation of the halo
major axis. The orientations of the second and third principal axis
are not documented, so we marginalize over them as described below. We
used the \texttt{axisratio\_[23]\_1} attributes to quantify the axis
ratios. For a given halo, we first compute the angle $\psi$ between
the halo's major axis and the CSV given by the halo-pair catalog; see
Figure~\ref{fig-ellipsoid}. We then calculate the skewer lengths
(hence $\frac{\Sigma_{\rm los}}{\Sigma_{\rm CSV}}$) for a variety of
LOS, and finally group the skewer results by the angle $\alpha$
between the CSV and the LOS.

\rev{The principal axes in the BigMDPL database are determined using
  the method of \citet{Allgood2006}, which starts with a spherical
  region of radius of $R_{\rm vir}$ and calculates a reduced
  inertia tensor considering the particles within this region. Then,
  they update the region to an ellipsoidal region using the obtained
  axis ratios, keeping the semimajor axis fixed at $R_{\rm vir}$,
  and redo the calculation, iterating this process until it converges.
  This is a widely adopted robust way to determine axis ratios, and
  \citet{Bett12} showed that the median axis ratios produced this way
  are similar (within 0.05) to those yielded by a simple inertia
  tensor. We found that the median $\frac{c}{a}$ for BigMDPL halos
  above $0.6\times10^{14} h^{-1} M_\odot$ is comparable to but
  slightly lower than those found in other studies: 0.46
  vs. 0.45--0.50 for a similar mass range in \citet{Tenneti15} and
  0.55 for the `reduced iterated' method in \citet{Bett12}.  The
  greater elongation of BigMDPL halos implies that we would find
  slightly smaller LOS bias effects in these other simulations.
  Still, it is not clear that BigMDPL halos are unduly elongated for
  our purposes given that halos are more elongated at higher density
  thresholds \citep{Jing2002} where DM scattering effects will be most
  important. Our median $\frac{c}{a}$ of 0.46 agrees well with the
  median $\frac{c}{a}$ found in \citet{Jing2002} for the high density
  threshold corresponding to the physical radius of 150 kpc we adopt
  below for averaging the surface mass density, even though the
  BigMDPL shapes were not explicitly evaluated at these high
  densities.}


\rev{We checked the reliability of these axis ratios in a few
  ways. First, the BigMDPL database includes a second set of axis
  ratios, \texttt{axisratio\_[23]\_1\_500c}, which are measured at a
  smaller radius than $0.3 R_{\rm vir}$. We found that the
  statistics were quite similar, with median axis ratios differing
  only by $\approx0.01$. Second, we found that in BigMDPL binary
  mergers, the axis ratios for the lower-mass partners match those of
  isolated halos (median $\frac{c}{a}=0.46$), but the more massive
  ('main') partner is more elongated ($\frac{c}{a}=0.35$).  This
  difference cannot be explained by particle misattribution near
  pericenter, which would presumably affect the lower-mass halo more.
  At least part of this difference may be the known effect that
  higher-mass halos are more elongated.}

\rev{\subsection{Weak Lensing}}\label{ssec-WL}

\begin{figure}
\centerline{\includegraphics[width=\columnwidth]{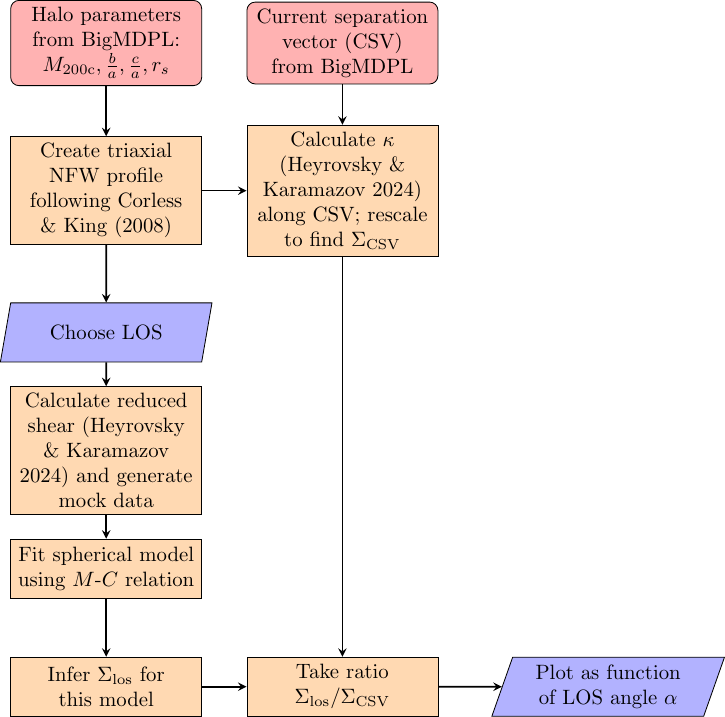}}
\caption{\rev{Summary of the workflow for establishing a triaxial
    model halo and extracting inferences from mock observations.}}
\label{fig-workflow}
\end{figure}

\rev{Inferences made via weak lensing average over a patch of sky and
  thus, unlike the central skewers, depend on the density
  profile. However, Figure~\ref{fig-sigmaskewer} shows little
  difference between profiles commonly used to fit clusters. We
  therefore adopt the more commonly used NFW profile. We use the
  cataloged BigMDPL halo parameters \texttt{m200c, rs,
    axisratio\_2\_1}, and \texttt{axisratio\_3\_1} to create a
  triaxial NFW model and fit it as follows. The workflow is summarized
  in Figure~\ref{fig-workflow}.  Following \citet{CorlessKing2008},
  we use the spherical scale radius \texttt{rs} and the axis ratios to
  define a triaxial scale radius $R_s$, then use their Equation 7 to
  recover the triaxial $R_{200}$ given \texttt{m200c}. The ratio
  $\frac{R_{200}}{R_s}$ then forms a triaxial concentration which
  yields a scale density $\rho_s$.  We then use the formalism of
  \citet{Heyrovsky24} to obtain the true $\bar{\Sigma}_{\rm CSV}$ of
  the triaxial halo, which we store for reference.  For this
  calculation, we use a fixed physical radius of 150 kpc as was used
  in the seminal dark matter constraint of \citet{Markevitch04},
  although we vary this radius later to probe the robustness of the
  results.}

\rev{We then choose a LOS and use the expressions in
  \citet{Heyrovsky24} to generate a reduced shear field for that
  LOS. We discard any ``observations'' of shear in a 30\arcs\ radius
  area around the center of the cluster, where
  cluster galaxies would prevent the detection of many background
  galaxies, and where the weak lensing approximation breaks down for
  massive clusters.  To set the size of the simulated field, we
  tested a series of sizes with spherical halos and found that at
  20\arcm\ and larger diameter, the fitted mass converges to the true
  mass. We opted to use a 30\arcm\ diameter field for the main run over all
  halos and all LOS.}

\rev{To simulate the process of estimating $\bar{\Sigma}$ as it would
  typically be done in an observational paper, we then fit the mock
  shear field with a \textit{spherical} NFW model following the
  \citet{Child2018} mass-concentration ($M$-$C$) relation, as weak lensing data are
  typically not informative enough to infer concentration or shape
  parameters. We fit with two free parameters, $M$ and $C$, which are
  linked by the $M$-$C$ relation but allowed to vary according to the
  scatter found by \citet{Child2018}.  In order to focus on the
  specific bias we are highlighting, we did not explore the effect of
  adding noise to the shear field. In that context, the central sky
  coordinates of the NFW halo were always well fit when allowed to
  float, so for efficiency in the main run we fixed those coordinates,
  leaving $M$ and $C$ as the only two free parameters.}

\rev{For each halo and LOS, we use the best-fit spherical model to
  infer $\bar{\Sigma}_{\rm los}$ over a 150 kpc radius patch and
  normalize it by the known $\bar{\Sigma}_{\rm CSV}$ for that halo.
  In \S\ref{sec-results} we explore how
  $\frac{\bar{\Sigma}_{\rm los}}{\bar{\Sigma}_{\rm CSV}}$ varies as a
  function of viewing angle and halo properties.}

\section{Results}\label{sec-results}

\subsection{Halo-merger alignment}

\begin{figure}
\centerline{\includegraphics[width=\columnwidth]{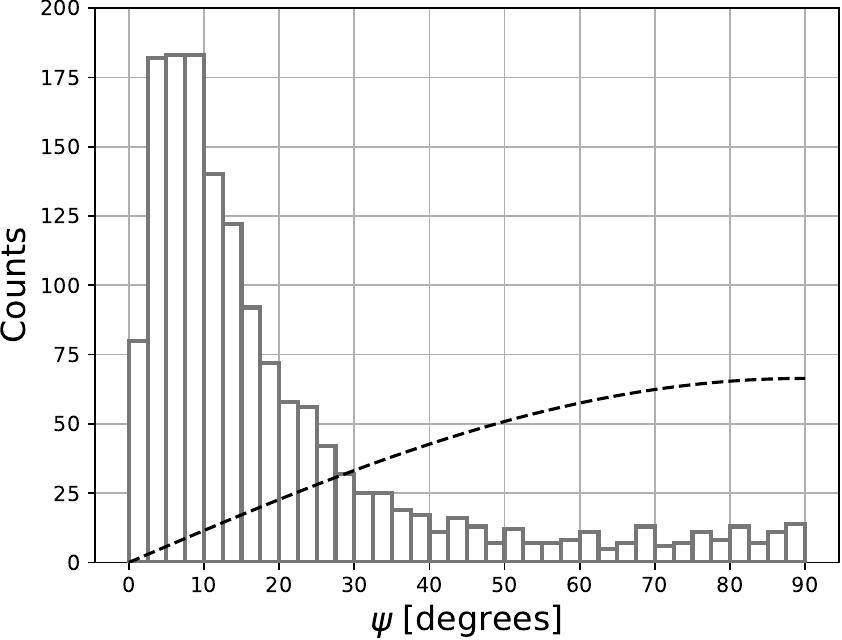}}
\caption{Histogram of the angle $\psi$ between the merger axis and
  each halo's major axis. The two axes show a remarkable tendency for
  alignment (small $\psi$), in contrast to the dashed curve which
  shows the distribution expected for $\psi$ distributed randomly on a
  sphere.}
\label{fig-HM}
\end{figure}

Figure~\ref{fig-HM} shows the aligment between the CSV and each halo's
major axis, in terms of a histogram of the angle $\psi$. The two axes
show a remarkable tendency for alignment (small $\psi$), in contrast
to the dashed curve which shows the distribution expected for CSVs
randomly distributed on a sphere. Assuming the clusters fell toward
each other along a connecting filament, this is consistent with work
finding that cluster-scale halos are aligned along larger-scale
filaments, which aligns them with neighboring halos
\citep{Kasun_2005}.

For binary merging clusters specifically, we would expect good
halo-halo alignment because both halos are falling toward each other
along a filament with which they are each aligned. We are not aware of
empirical measurements of this effect, which may be difficult given
overlapping halos. However, halo-halo alignment in binary mergers can
be inferred from two empirical results: the orientation of a brightest
cluster galaxy (BCG) is a proxy for the orientation of its host
cluster \citep{West17}, and BCGs in merging clusters are known to
align with the CSV \citep{WittmanFooteBCGalignment}.  \rev{Although we
  do not attempt a direct comparison because those works involve 2-D
  measurements of galaxies rather than 3-D measurements of clusters,
  the alignment seen in Figure~\ref{fig-HM} is about as strong as seen
  in those works.}

\subsection{Skewer surface mass density}\label{ssec-skewer}

\begin{figure}
  \centerline{\includegraphics[width=\columnwidth]{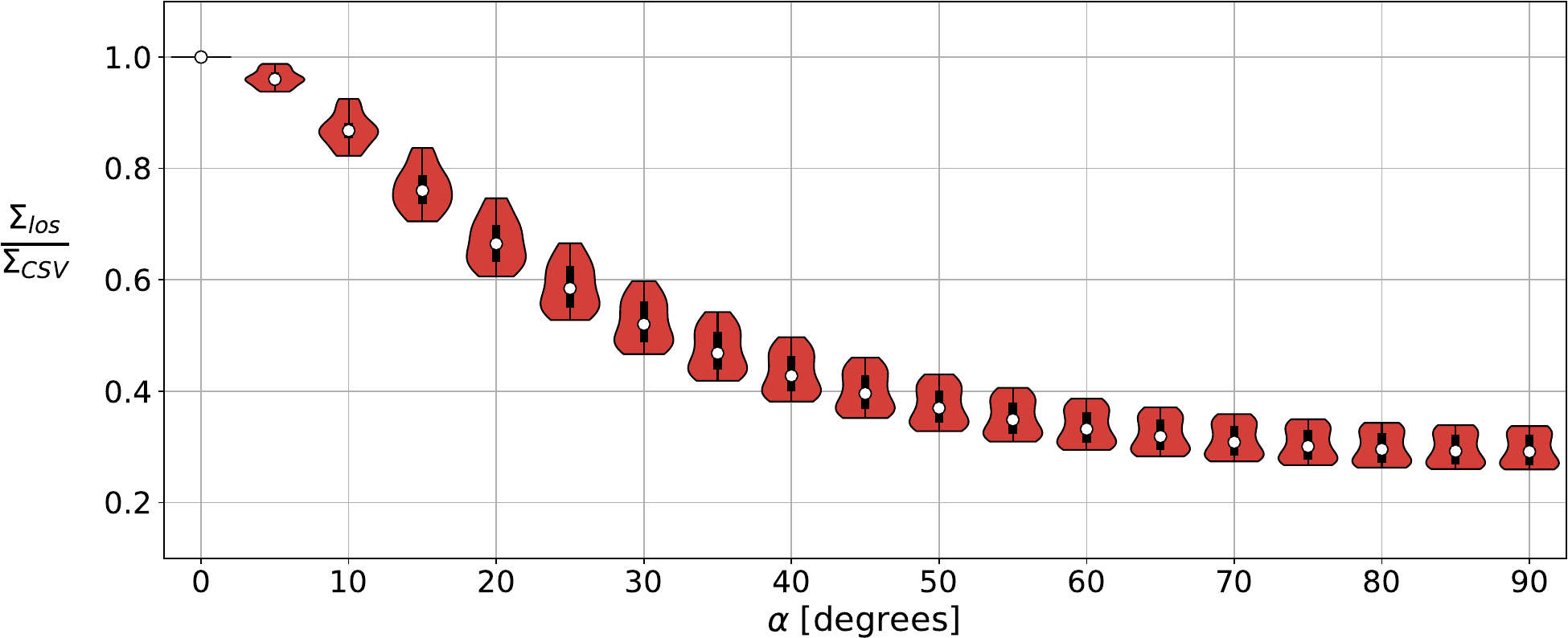}}\centerline{\includegraphics[width=\columnwidth]{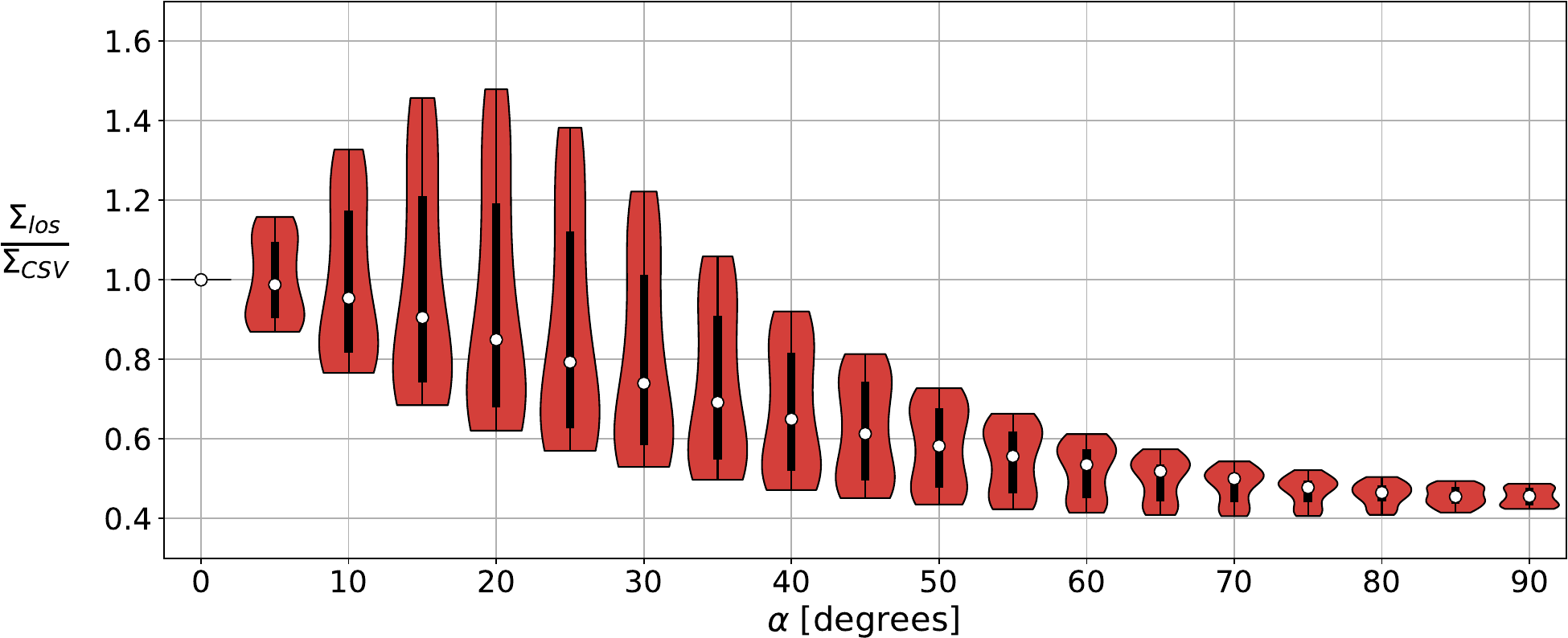}}
  \caption{Skewer bias factor $\frac{\Sigma_{\rm los}}{\Sigma_{\rm CSV}}$ as
    a function of viewing angle $\alpha$ for two halos with similar
    axis ratios. Top:
    halo 13410052637 is well aligned with its merger axis
    ($\psi=1.5^\circ$), such that lines of sight perpendicular to the
    merger axis (large $\alpha$) encounter systematically low column
    density. Bottom: halo 13582533817 is less well aligned
    ($\psi=18^\circ$), leading to more mixed results but still with
    bias at large $\alpha$.  In each case the violin shows the
    distribution, which the open circle marks the median and the black
    band marks the middle quartiles.}
\label{fig-exhalos}
\end{figure}

\textit{Illustrative halos.} Figure~\ref{fig-exhalos}, top panel,
\rev{illustrates how $\frac{\Sigma_{\rm los}}{\Sigma_{\rm CSV}}$
  varies with $\alpha$ for a very well aligned and very elongated}
halo ($\psi=1.5^\circ$, $\frac{b}{a}=0.337$,
$\frac{c}{a}=0.259$). Results are shown as violin plots for values of
$\alpha$ in 5$^\circ$ increments. At $\alpha=0$,
$\frac{\Sigma_{\rm los}}{\Sigma_{\rm CSV}}=1$ by definition and only
one LOS is sampled so no violin appears. As $\alpha$ increases, the
LOS pierces the halo closer to the intermediate or minor axis, so
$\frac{\Sigma_{\rm los}}{\Sigma_{\rm CSV}}$ falls below
unity. Furthermore, at each $\alpha>0$ there are multiple LOS,
yielding a range of $\frac{\Sigma_{\rm los}}{\Sigma_{\rm CSV}}$
depending on their proximity to the halo minor or intermediate axis.
Observed systems are typically identified as mergers and used for DM
constraints only if $\alpha>45^\circ$, so this halo suggests that
$\frac{\Sigma_{\rm los}}{\Sigma_{\rm CSV}}\approx 0.3$. In other
words, the surface mass density encountered by DM particles through
the pericenter passage \rev{in this rather extreme case} is about
three times that measured along the LOS, so the resultant DM
constraint would be biased high by a factor of three.

Figure~\ref{fig-exhalos}, bottom panel, shows an \rev{equally
  elongated halo with a more typical} $\psi=18^\circ$ misalignment
from the CSV.  Therefore the maximum $\Sigma_{\rm los}$ is encountered
by one LOS at $\alpha=18^\circ$: the LOS that happens to align with
the major axis. Other LOS at $\alpha=18^\circ$ encounter a range of
lower column densities. As one proceeds to higher $\alpha$ the LOS are
sampling the intermediate and minor axes---but to a lesser degree than
with a well-aligned halo, hence $\Sigma_{\rm los}$ does not drop as
far. \rev{As a reminder, these are highly elongated halos chosen to
  illustrate the dependence of
  $\frac{\Sigma_{\rm los}}{\Sigma_{\rm CSV}}$ on halo alignment $\psi$
  and viewing angle $\alpha$.  We now proceed to marginalize over all
  merging halos.}

\begin{figure}
  \centerline{\includegraphics[width=\columnwidth]{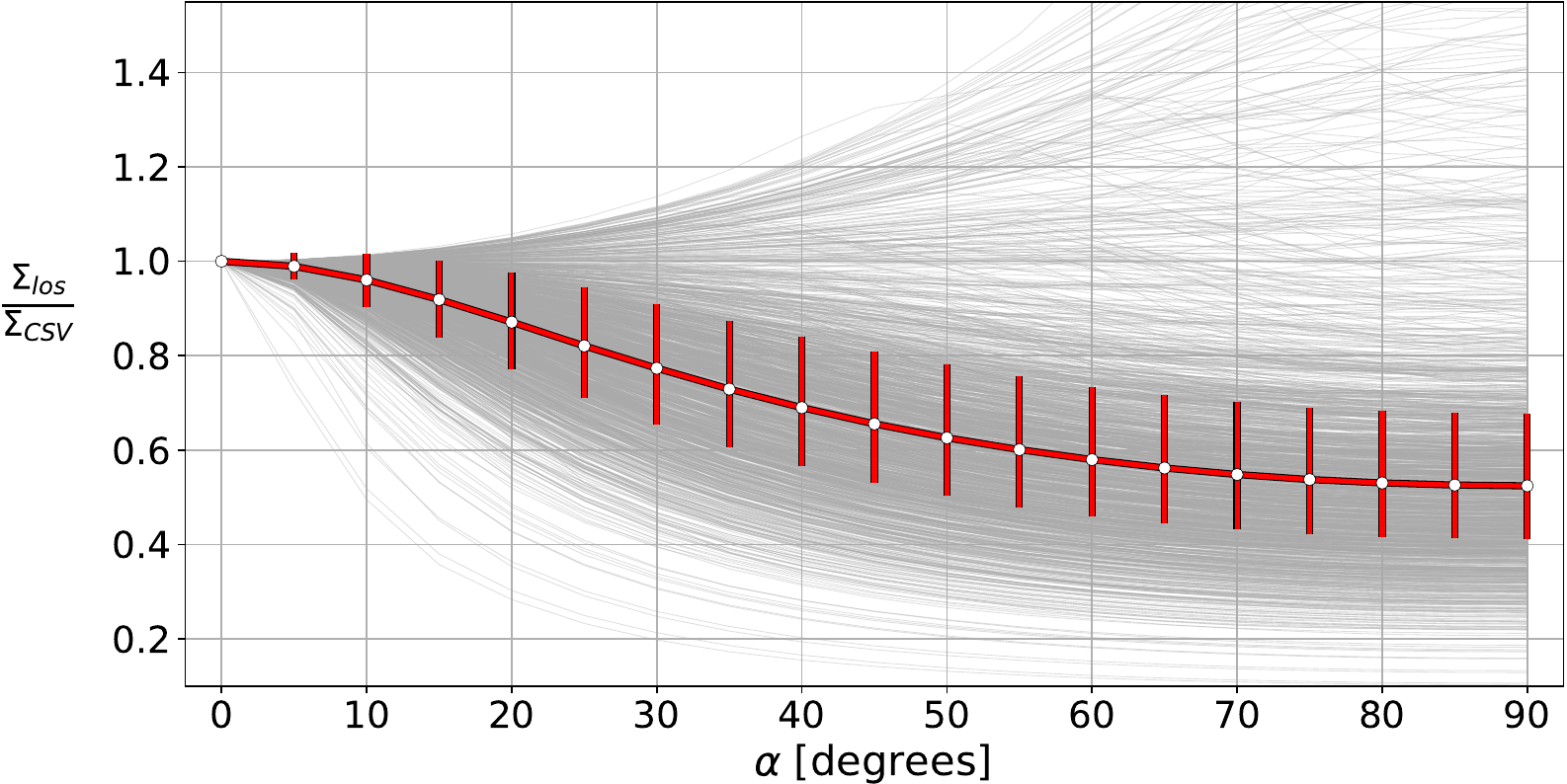}}
  \caption{Median (circles) and interquartile range (bars) of the
    skewer bias factor marginalized over all halos, as a function of
    viewing angle $\alpha$.  For typical observed cases
    $\alpha>45^\circ$ so the \rev{central} surface mass density along
    the LOS is about half that encountered by DM particles along the
    separation vector. \rev{The gray curves illustrate the median behavior of
      each individual halo (medianized over various LOS at fixed $\alpha$).}}
\label{fig-allhalos}
\end{figure}

\textit{All halos.} A few halo pairs provided remarkably large or
small bias factors, for example when one halo's major axis points at
the other's minor axis in a ``T'' shape. We suspect that these extreme
cases are artifacts of the halo finder having difficulty separating
overlapping halos. To reduce clutter from these possibly unphysical
cases, in Figure~\ref{fig-allhalos} we show the median and
interquartile range (IQR) of the bias factor distribution after
marginalizing over all halos in the relevant halo pairs. Again,
considering that observed cases will have $\alpha>45^\circ$, the
surface mass density encountered by DM particles through the
pericenter passage is nearly twice that measured along the LOS.
\rev{Given that the central skewer estimates are a lower limit on
  $\frac{\bar{\Sigma}_{\rm los}}{\bar{\Sigma}_{\rm CSV}}$, we expect
  that resulting DM constraints would be biased high, but by less than
  a factor of two.}

\begin{figure}
\centerline{\includegraphics[width=\columnwidth]{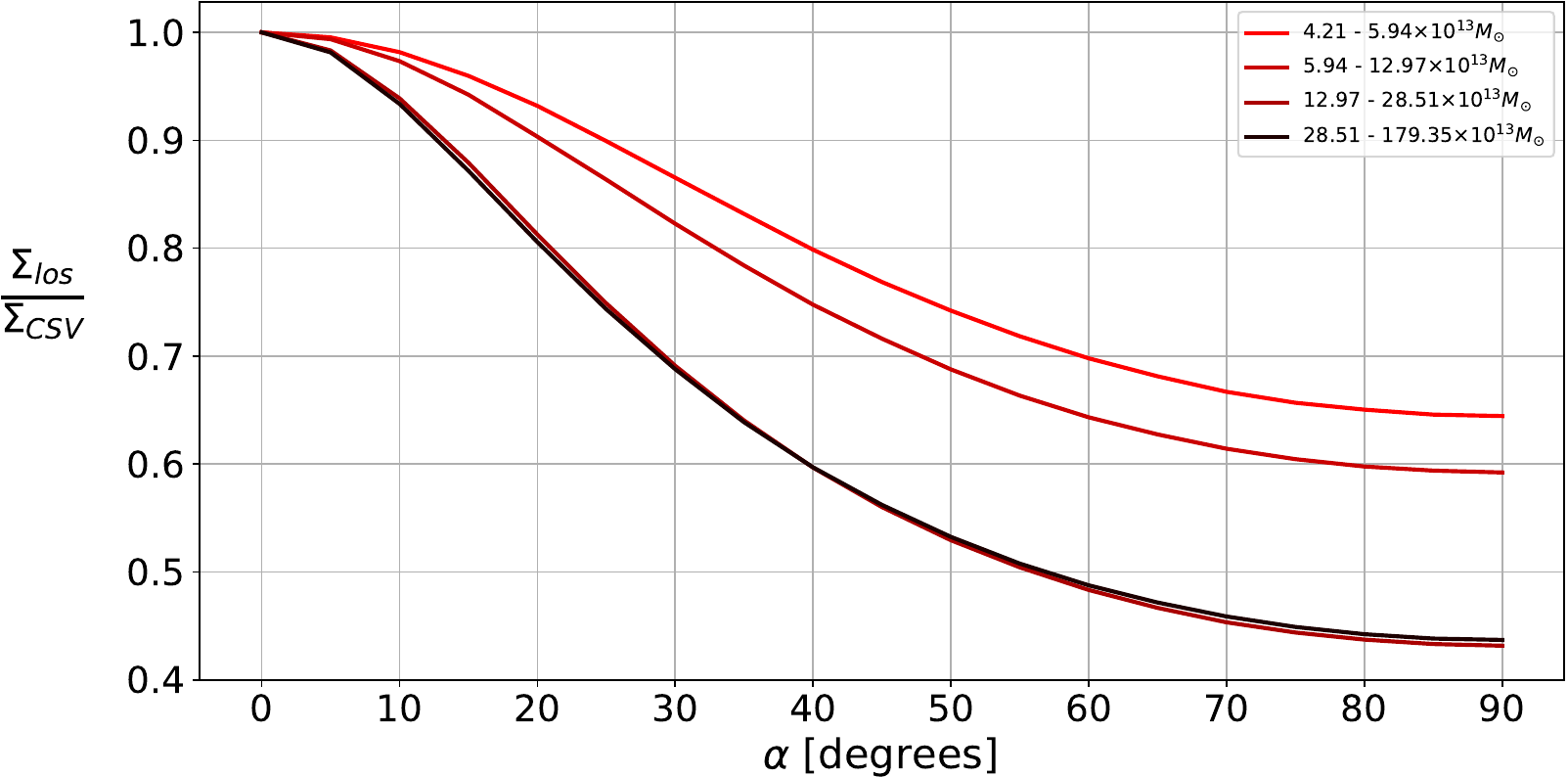}}
\caption{Median skewer bias factor $\frac{\Sigma_{\rm los}}{\Sigma_{\rm CSV}}$ as a
  function of viewing angle $\alpha$, marginalized over all halos in a
  given mass quartile. Lower-mass halos show reduced LOS effects
  because they are rounder.
\label{fig-skewermassquartiles}}
\end{figure}

\textit{Effect of mass.} Figure~\ref{fig-skewermassquartiles} shows
how these trends differ across halo mass quartiles; for clarity only
the median effect at each $\alpha$ and mass bin is shown. Lower-mass
halos are rounder \citep[in BigMDPL as well as in other works,
e.g.][]{Allgood06,Henson16} so the effects of LOS are less marked at
lower mass.

\subsection{Mock weak lensing results} 

\begin{figure}
  \centerline{\includegraphics[width=\columnwidth]{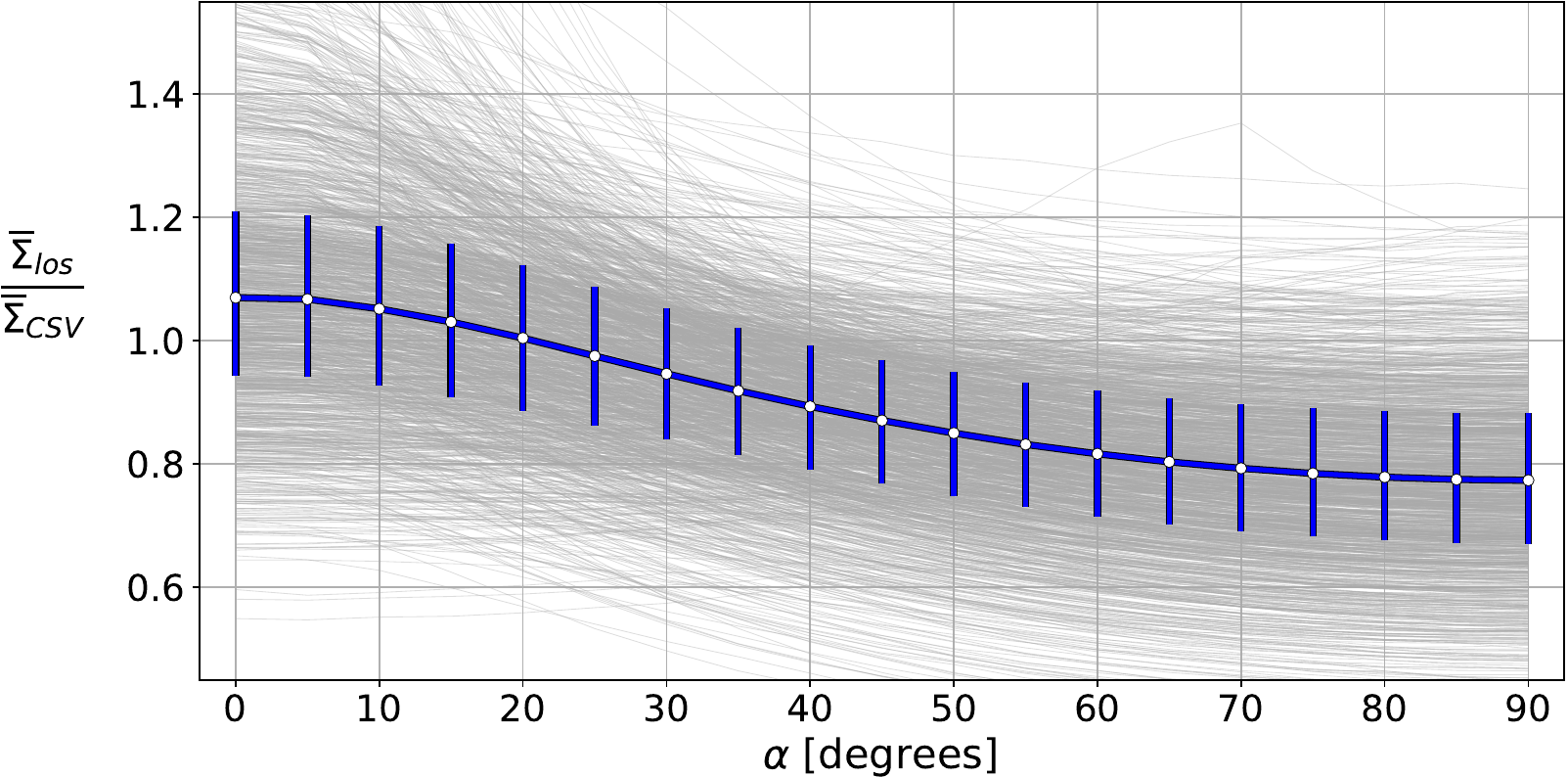}}
  \caption{Weak lensing bias factor as a function of viewing angle
    $\alpha$, marginalized over all halos. The circles indicate
    medians and the bars indicate the interquartile range. \rev{The
      gray curves illustrate the median behavior of each individual
      halo (medianized over various LOS at fixed $\alpha$).} The trend
    is \rev{similar to that seen for the skewer estimate}, but the
    bias factor is less extreme.}
\label{fig-WL}
\end{figure}

\rev{We show the WL results as a function of $\alpha$ in
  Figure~\ref{fig-WL}. Each halo is shown as a gray curve, with the
  medians and interquartile ranges overlaid. As expected, the results
  are not as extreme as the central skewer calculation: at large
  $\alpha$ the bias is about 0.8.  This LOS bias happens to be of
    similar magnitude to that found by \citet{Giocoli2024} for total
    mass.} They performed mock WL fits along the principal axes of
  clusters in the Three Hundred simulation \citep{Cui2018} and found
  that orientations along the (major, intermediate, minor) axes bias
  the inference of the true mass by factors of about (1.2, 0.85,
  0.75).\footnote{The Three Hundred halos include baryons so should be
    slightly rounder than our BigMDPL halos which are dark matter
    only. However, this is a small effect; see
    \S\ref{sec-discussion}.} \rev{However, this is a numerical
    coincidence because our calculation is for surface mass density
    within 150 kpc projected radius rather than mass, and we show
    below that the result is sensitive to that particular choice.}

\rev{Note one distinction between the construction of the skewer and
  WL plots: for the skewers, the ratio
  $\frac{\Sigma_{\rm los}}{\Sigma_{\rm CSV}}$ was \textit{defined} to
  be unity at $\alpha=0$. For the WL plots, observational effects and
  modeling choices could cause the inferred $\bar{\Sigma}_{\rm los}$
  to differ from the true $\bar{\Sigma}_{\rm CSV}$ even when the LOS
  is along the CSV. (In fact, Figure~\ref{fig-WL} shows that the
  median ratio is slightly above unity at $\alpha=0$.)  This choice in
  plot construction also forces the various halo curves to meet at
  $\alpha=0$ in the skewer plot (Figure~\ref{fig-allhalos}) but not
  the WL plot (Figure~\ref{fig-WL}). This causes an apparent variation
  in halo-to-halo scatter as a function of $\alpha$ in
  Figure~\ref{fig-allhalos}.}

\begin{figure}
  \centerline{\includegraphics[width=\columnwidth]{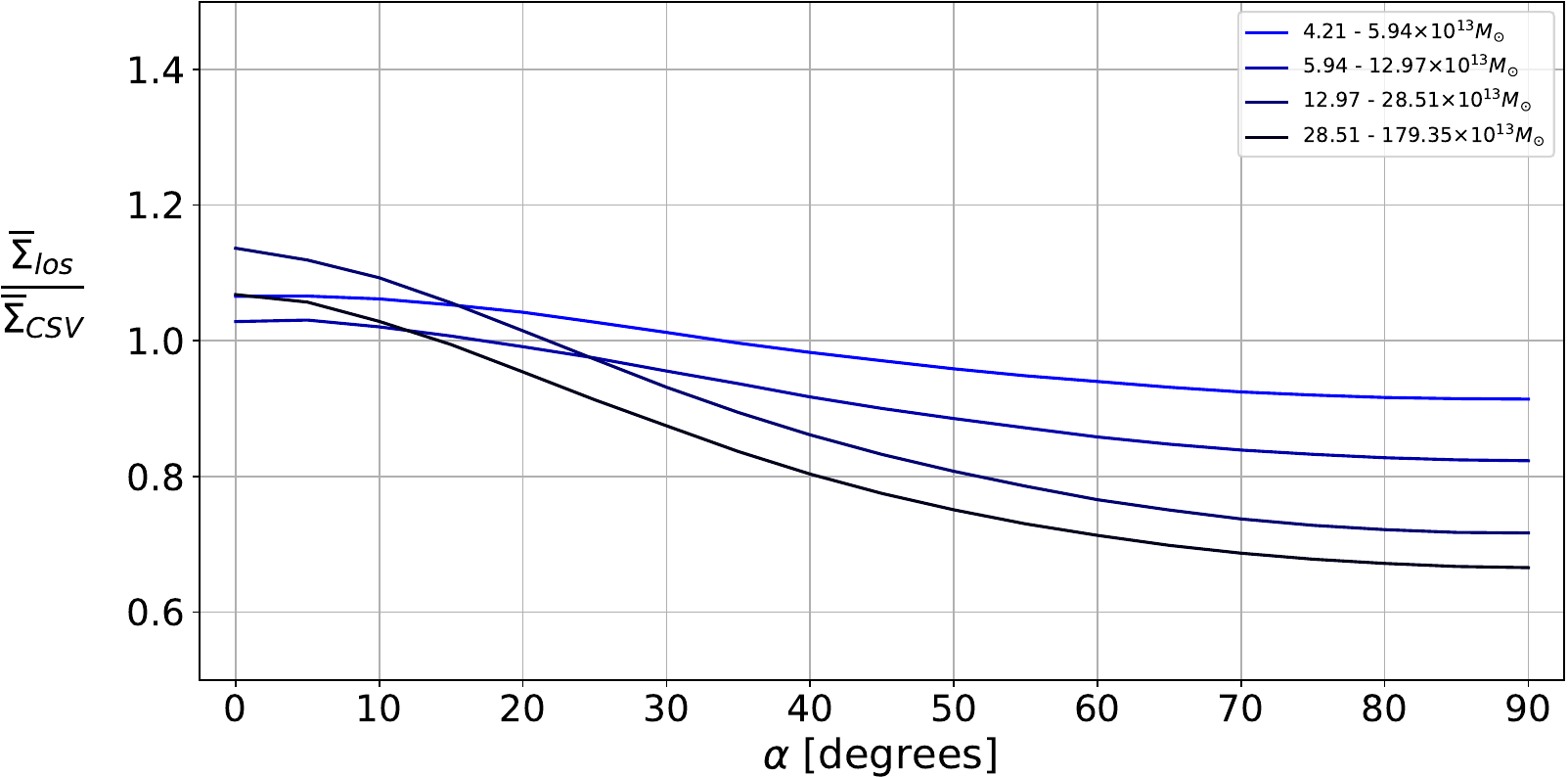}}
\caption{Median weak lensing bias factor as a function
  of viewing angle $\alpha$, marginalized over all halos in a given
  mass quartile. }
\label{fig-WL2}
\end{figure}

Figure~\ref{fig-WL2} shows the same results split by mass 
quartile. \rev{As expected, we find that the higher-mass halos are 
  most sensitive to LOS effects, because they have more extreme axis ratios.}

\begin{figure}
  \centerline{\includegraphics[width=\columnwidth]{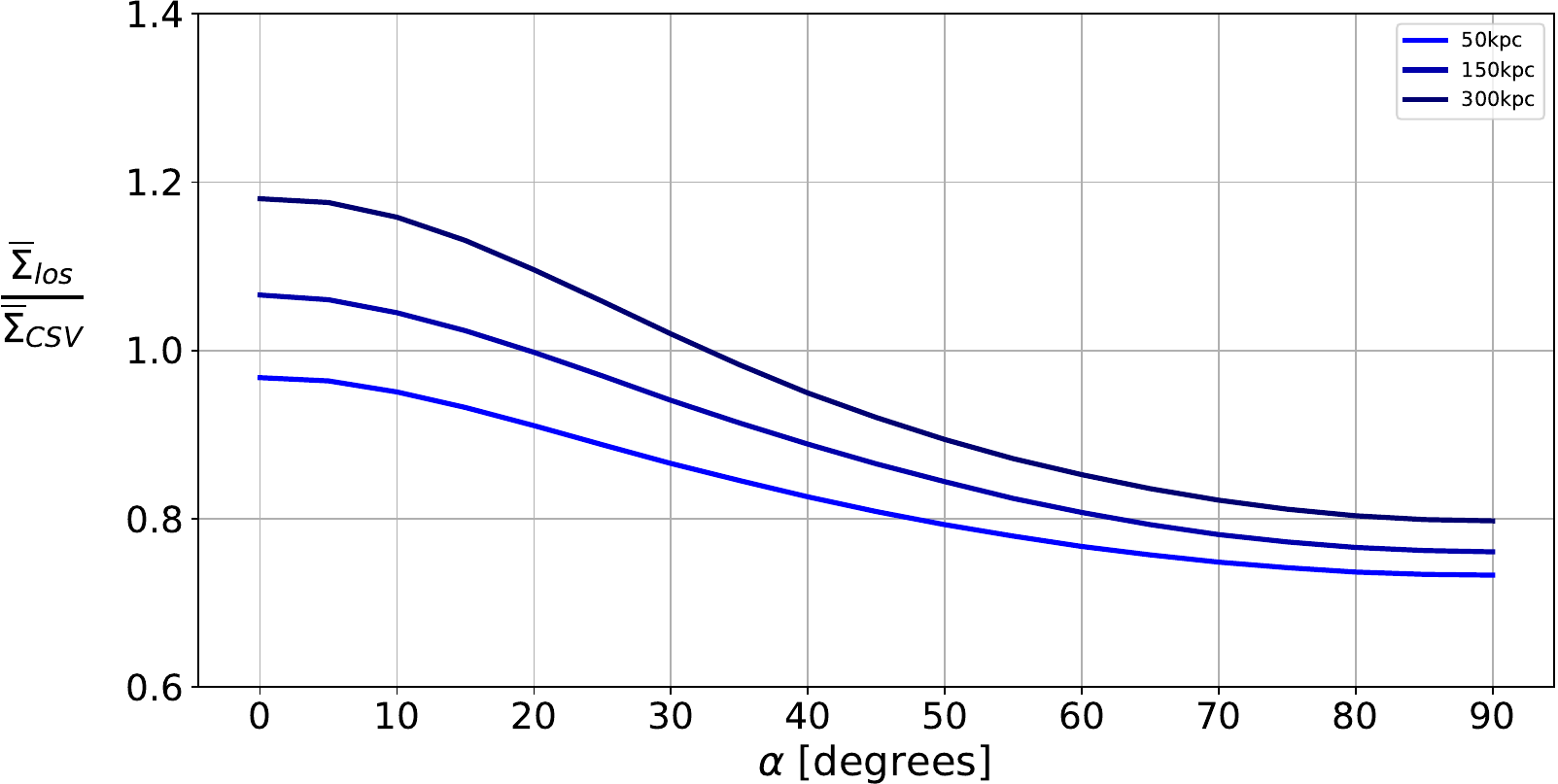}}
  \caption{\rev{Our fiducial weak lensing bias factor was defined in terms
    of surface mass density projected within 150 kpc of the halo
    center. Here we show how the bias factor depends on this choice.}}
\label{fig-Rvary}
\end{figure}

To this point, the weak lensing bias factor has been defined in terms
of surface mass density projected within 150 kpc of the halo
center. Figure~\ref{fig-Rvary} shows the median bias (marginalized
over all halos) as a function of $\alpha$ for two other choices of
radius, 50 and 300 kpc. There is substantial variation, but mostly at
small $\alpha$ which will rarely apply to observations in practice. At
large $\alpha$, the variation is about $\pm0.05$. This is comparable
to the variation that might be seen with the use of different halo
catalogs as discussed in \S\ref{sec-m-simulation}.

\section{Summary and discussion}\label{sec-discussion}

We have found that current \sDM\ constraints based on the scattering
depth argument are on average biased because spherical symmetry is
violated: the surface mass density along the LOS is lower than that
experienced by DM particles as they pass along the merger axis.
Because the surface mass density is inversely related to the inferred
cross section, this yields an overestimate of the DM cross section.
\rev{We cast our results in terms of
  $\frac{\bar{\Sigma}_{\rm los}}{\bar{\Sigma}_{\rm CSV}}$, which in
  this argument is the same as the ratio of true to inferred DM cross
  section. We first introduced a thinking tool---skewers running
  through triaxial halo centers---that provides a strict lower bound
  to this ratio, independent of the density profile. In
  \S\ref{ssec-skewer} we found that this ratio had a median value of
  0.5 for typically used LOS, and with large scatter from halo to
  halo.  Our weak lensing simulations (\S\ref{ssec-WL}) show that in
  practice the ratio is about 0.8 for typically used LOS, and with
  less scatter. It also shows a modest dependence on the size of
  the region over which the surface mass density is averaged.} Note
that this bias is distinct from the overall mass bias studied in
\citet{WonkiBias2023}, which is smaller and stems from the overlap of
two hypothetically spherical halos.

We studied dark matter only (DMO) halos; the presence of baryons will
reduce bias by making halos rounder, but only slightly. For example,
the simulations of \citet{Henson16} found that the median minor/major axis
ratio shifts from $0.537$ for DMO to $0.576$ with baryons. This is
only 40\% of the scatter across halos. The intermediate/major axis
ratio shifts by a similarly small amount when baryons are added.
\rev{Still, there remains some uncertainty regarding the size of the
  effect given that different simulations with the same physics have
  different median axis ratios: the axis ratios for our BigMDPL halos
  are nontrivially lower than the DMO value in \citet{Henson16}. A
  comparable uncertainty is due to the size of the region within
  which $\bar{\Sigma}$ is measured (Figure~\ref{fig-Rvary}), so further
  study is warranted---although some of these uncertainties will be
  obviated when DM constraints are derived from simulations as
  described in our final paragraph.}

The bias is a function of halo and LOS properties, so more precise
estimates could in principle be tailored to each individual observed
system.  Such tailoring would involve a straightforward extension of
the procedure here: rather than counting all halo pairs with equal
weight in the marginalization, one should weight by the likelihood of
a halo pair matching the observables (subhalo projected separation,
mass, and line-of-sight relative velocity).

Most known binary systems were identified due to their clear separation
between subclusters on the sky, hence $\alpha$ is not small. This is
why we have referred to lines of sight at large $\alpha$ as
``typical.'' This is likely to remain the case for any cluster
amenable to this DM scattering argument; for mergers along
the LOS there is no mechanism for measuring a DM offset.

Although our argument was framed in terms of analytical
 approximations relating \sDM\ to scattering depth
 \citep{Markevitch04,Harvey14}, the same bias applies to staged
 simulations based on the assumption of spherical symmetry. To
 overcome this bias, staged simulations should be staged with
 appropriate halo shapes and alignments, and we urge simulators to do
 so.  \rev{We checked the time evolution of the alignment, and we
   found that main halos are already well aligned at least ${\sim}1$
   Gyr before pericenter. Hence one cannot rely on the alignment
   arising organically from processes (e.g. tidal effects) already
   present in the staged simulation.}  A further step would be to go
 beyond the triaxial approximation and resimulate merging systems
 found in cosmological simulations.  This would incorporate the effect
 of finer substructure and the fact that axis ratios vary with density
 \citep{Jing2002}.  \rev{Such} simulations would also naturally
 incorporate the appropriate range of halo alignments and pericenter
 distances.


\acknowledgments We thank the anonymous referee for numerous
constructive suggestions. This work was supported by NSF grant number
2308383.  The CosmoSim database used in this paper is a service by the
Leibniz-Institute for Astrophysics Potsdam (AIP). The MultiDark
database was developed in cooperation with the Spanish MultiDark
Consolider Project CSD2009-00064.  The authors gratefully acknowledge
the Gauss Centre for Supercomputing e.V. (www.gauss-centre.eu) and the
Partnership for Advanced Supercomputing in Europe (PRACE,
www.prace-ri.eu) for funding the MultiDark simulation project by
providing computing time on the GCS Supercomputer SuperMUC at Leibniz
Supercomputing Centre (LRZ, www.lrz.de).



\bibliography{ms}

\begin{thebibliography}{}
\expandafter\ifx\csname natexlab\endcsname\relax\def\natexlab#1{#1}\fi

\bibitem[{Allgood {et~al.}(2006)Allgood, Flores, Primack, Kravtsov, Wechsler,
  Faltenbacher, \& Bullock}]{Allgood2006}
Allgood, B., Flores, R.~A., Primack, J.~R., {et~al.} 2006, Monthly Notices of
  the Royal Astronomical Society, 367, 1781

\bibitem[{{Allgood} {et~al.}(2006){Allgood}, {Flores}, {Primack}, {Kravtsov},
  {Wechsler}, {Faltenbacher}, \& {Bullock}}]{Allgood06}
{Allgood}, B., {Flores}, R.~A., {Primack}, J.~R., {et~al.} 2006, \mnras, 367,
  1781

\bibitem[{{Behroozi} {et~al.}(2013){Behroozi}, {Wechsler}, \&
  {Wu}}]{Rockstar2013}
{Behroozi}, P.~S., {Wechsler}, R.~H., \& {Wu}, H.-Y. 2013, \apj, 762, 109

\bibitem[{Bett(2012)}]{Bett12}
Bett, P. 2012, Monthly Notices of the Royal Astronomical Society, 420, 3303

\bibitem[{{Binggeli}(1982)}]{Binggeli1982}
{Binggeli}, B. 1982, \aap, 107, 338

\bibitem[{{Child} {et~al.}(2018){Child}, {Habib}, {Heitmann}, {Frontiere},
  {Finkel}, {Pope}, \& {Morozov}}]{Child2018}
{Child}, H.~L., {Habib}, S., {Heitmann}, K., {et~al.} 2018, \apj, 859, 55

\bibitem[{{Corless} \& {King}(2008)}]{CorlessKing2008}
{Corless}, V.~L., \& {King}, L.~J. 2008, \mnras, 390, 997

\bibitem[{{Cui} {et~al.}(2018){Cui}, {Knebe}, {Yepes}, {Pearce}, {Power},
  {Dave}, {Arth}, {Borgani}, {Dolag}, {Elahi}, {Mostoghiu}, {Murante}, {Rasia},
  {Stoppacher}, {Vega-Ferrero}, {Wang}, {Yang}, {Benson}, {Cora}, {Croton},
  {Sinha}, {Stevens}, {Vega-Mart{\'\i}nez}, {Arthur}, {Baldi}, {Ca{\~n}as},
  {Cialone}, {Cunnama}, {De Petris}, {Durando}, {Ettori}, {Gottl{\"o}ber},
  {Nuza}, {Old}, {Pilipenko}, {Sorce}, \& {Welker}}]{Cui2018}
{Cui}, W., {Knebe}, A., {Yepes}, G., {et~al.} 2018, \mnras, 480, 2898

\bibitem[{{Dawson} {et~al.}(2011){Dawson}, {Wittman}, {Jee}, {Gee}, {Hughes},
  {Tyson}, {Schmidt}, {Thorman}, {Bradac}, {Miyazaki}, {Lemaux}, \&
  {Utsumi}}]{Dawson11}
{Dawson}, W.~A., {Wittman}, D., {Jee}, M., {et~al.} 2011, ArXiv e-prints,
  arXiv:1110.4391

\bibitem[{{Einasto}(1965)}]{Einasto1965}
{Einasto}, J. 1965, Trudy Astrofizicheskogo Instituta Alma-Ata, 5, 87

\bibitem[{{Euclid Collaboration} {et~al.}(2024){Euclid Collaboration},
  {Giocoli}, {Meneghetti}, {Rasia}, {Borgani}, {Despali}, {Lesci}, {Marulli},
  {Moscardini}, {Sereno}, {Cui}, {Knebe}, {Yepes}, {Castro}, {Corasaniti},
  {Pires}, {Castignani}, {Schrabback}, {Pratt}, {Le Brun}, {Aghanim},
  {Amendola}, {Auricchio}, {Baldi}, {Bodendorf}, {Bonino}, {Branchini},
  {Brescia}, {Brinchmann}, {Camera}, {Capobianco}, {Carbone}, {Carretero},
  {Castander}, {Castellano}, {Cavuoti}, {Cledassou}, {Congedo}, {Conselice},
  {Conversi}, {Copin}, {Corcione}, {Courbin}, {Cropper}, {Da Silva},
  {Degaudenzi}, {Dinis}, {Dubath}, {Dupac}, {Dusini}, {Farrens}, {Ferriol},
  {Fosalba}, {Frailis}, {Franceschi}, {Fumana}, {Galeotta}, {Garilli},
  {Gillis}, {Grazian}, {Grupp}, {Haugan}, {Holmes}, {Hornstrup}, {Jahnke},
  {K{\"u}mmel}, {Kermiche}, {Kilbinger}, {Kunz}, {Kurki-Suonio}, {Ligori},
  {Lilje}, {Lloro}, {Maiorano}, {Mansutti}, {Marggraf}, {Markovic}, {Massey},
  {Maurogordato}, {Mei}, {Merlin}, {Meylan}, {Moresco}, {Munari}, {Niemi},
  {Nightingale}, {Nutma}, {Padilla}, {Paltani}, {Pasian}, {Pedersen},
  {Pettorino}, {Polenta}, {Poncet}, {Popa}, {Raison}, {Renzi}, {Rhodes},
  {Riccio}, {Romelli}, {Roncarelli}, {Rossetti}, {Saglia}, {Sapone},
  {Sartoris}, {Schneider}, {Secroun}, {Serrano}, {Sirignano}, {Sirri},
  {Stanco}, {Starck}, {Tallada-Cresp{\'\i}}, {Taylor}, {Tereno},
  {Toledo-Moreo}, {Torradeflot}, {Tutusaus}, {Valentijn}, {Valenziano},
  {Vassallo}, {Wang}, {Weller}, {Zamorani}, {Zoubian}, {Andreon}, {Bardelli},
  {Boucaud}, {Bozzo}, {Colodro-Conde}, {Di Ferdinando}, {Fabbian}, {Farina},
  {Israel}, {Keih{\"a}nen}, {Lindholm}, {Mauri}, {Neissner}, {Schirmer},
  {Scottez}, {Tenti}, {Zucca}, {Akrami}, {Baccigalupi}, {Ballardini},
  {Bernardeau}, {Biviano}, {Borlaff}, {Burigana}, {Cabanac}, {Cappi},
  {Carvalho}, {Casas}, {Chambers}, {Cooray}, {Courtois}, {Davini}, {de la
  Torre}, {De Lucia}, {Desprez}, {Dole}, {Escartin}, {Escoffier}, {Ferrero},
  {Finelli}, {Gabarra}, {Ganga}, {Garcia-Bellido}, {George}, {Giacomini},
  {Gozaliasl}, {Hildebrandt}, {Hook}, {Jimenez Mu{\~n}oz}, {Joachimi},
  {Kajava}, {Kansal}, {Kirkpatrick}, {Legrand}, {Loureiro}, {Macias-Perez},
  {Magliocchetti}, {Mainetti}, {Maoli}, {Marcin}, {Martinelli}, {Martinet},
  {Martins}, {Matthew}, {Maurin}, {Metcalf}, {Monaco}, {Morgante}, {Nadathur},
  {Nucita}, {Patrizii}, {Peel}, {Pollack}, {Popa}, {Porciani}, {Potter},
  {P{\"o}ntinen}, {Reimberg}, {S{\'a}nchez}, {Sakr}, {Schneider}, {Sefusatti},
  {Shulevski}, {Spurio Mancini}, {Stadel}, {Steinwagner}, {Valiviita},
  {Veropalumbo}, {Viel}, \& {Zinchenko}}]{Giocoli2024}
{Euclid Collaboration}, {Giocoli}, C., {Meneghetti}, M., {et~al.} 2024, \aap,
  681, A67

\bibitem[{{Harvey} {et~al.}(2015){Harvey}, {Massey}, {Kitching}, {Taylor}, \&
  {Tittley}}]{Harvey15}
{Harvey}, D., {Massey}, R., {Kitching}, T., {Taylor}, A., \& {Tittley}, E.
  2015, Science, 347, 1462

\bibitem[{{Harvey} {et~al.}(2014){Harvey}, {Tittley}, {Massey}, {Kitching},
  {Taylor}, {Pike}, {Kay}, {Lau}, \& {Nagai}}]{Harvey14}
{Harvey}, D., {Tittley}, E., {Massey}, R., {et~al.} 2014, \mnras, 441, 404

\bibitem[{Henson {et~al.}(2016)Henson, Barnes, Kay, McCarthy, \&
  Schaye}]{Henson16}
Henson, M.~A., Barnes, D.~J., Kay, S.~T., McCarthy, I.~G., \& Schaye, J. 2016,
  Monthly Notices of the Royal Astronomical Society, 465, 3361

\bibitem[{{Heyrovsk{\'y}} \& {Karamazov}(2024)}]{Heyrovsky24}
{Heyrovsk{\'y}}, D., \& {Karamazov}, M. 2024, arXiv e-prints, arXiv:2404.00169

\bibitem[{{Jing} \& {Suto}(2002)}]{Jing2002}
{Jing}, Y.~P., \& {Suto}, Y. 2002, \apj, 574, 538

\bibitem[{{Kaplinghat} {et~al.}(2016){Kaplinghat}, {Tulin}, \&
  {Yu}}]{Kaplinghat16}
{Kaplinghat}, M., {Tulin}, S., \& {Yu}, H.-B. 2016, \prl, 116, 041302

\bibitem[{Kasun \& Evrard(2005)}]{Kasun_2005}
Kasun, S.~F., \& Evrard, A.~E. 2005, The Astrophysical Journal, 629, 781

\bibitem[{{Klypin} {et~al.}(2016){Klypin}, {Yepes}, {Gottl{\"o}ber}, {Prada},
  \& {He{\ss}}}]{BigMDPL2016}
{Klypin}, A., {Yepes}, G., {Gottl{\"o}ber}, S., {Prada}, F., \& {He{\ss}}, S.
  2016, \mnras, 457, 4340

\bibitem[{{Lee} {et~al.}(2023){Lee}, {Cha}, {Jee}, {Nagai}, {King}, {ZuHone},
  {Chadayammuri}, {Felix}, \& {Finner}}]{WonkiBias2023}
{Lee}, W., {Cha}, S., {Jee}, M.~J., {et~al.} 2023, \apj, 945, 71

\bibitem[{{Markevitch} {et~al.}(2004){Markevitch}, {Gonzalez}, {Clowe},
  {Vikhlinin}, {Forman}, {Jones}, {Murray}, \& {Tucker}}]{Markevitch04}
{Markevitch}, M., {Gonzalez}, A.~H., {Clowe}, D., {et~al.} 2004, \apj, 606, 819

\bibitem[{{Navarro} {et~al.}(1997){Navarro}, {Frenk}, \& {White}}]{NFW97}
{Navarro}, J.~F., {Frenk}, C.~S., \& {White}, S.~D.~M. 1997, \apj, 490, 493

\bibitem[{{Plionis}(1994)}]{Plionis1994}
{Plionis}, M. 1994, \apjs, 95, 401

\bibitem[{{Randall} {et~al.}(2008){Randall}, {Markevitch}, {Clowe}, {Gonzalez},
  \& {Brada{\v c}}}]{Randall2008}
{Randall}, S.~W., {Markevitch}, M., {Clowe}, D., {Gonzalez}, A.~H., \&
  {Brada{\v c}}, M. 2008, ApJ, 679, 1173

\bibitem[{{Riebe} {et~al.}(2013){Riebe}, {Partl}, {Enke}, {Forero-Romero},
  {Gottl{\"o}ber}, {Klypin}, {Lemson}, {Prada}, {Primack}, {Steinmetz}, \&
  {Turchaninov}}]{CosmoSim2013}
{Riebe}, K., {Partl}, A.~M., {Enke}, H., {et~al.} 2013, Astronomische
  Nachrichten, 334, 691

\bibitem[{{Robertson} {et~al.}(2017){Robertson}, {Massey}, \&
  {Eke}}]{Robertson17Bullet}
{Robertson}, A., {Massey}, R., \& {Eke}, V. 2017, \mnras, 465, 569

\bibitem[{{Smargon} {et~al.}(2012){Smargon}, {Mandelbaum}, {Bahcall}, \&
  {Niederste-Ostholt}}]{Smargon2012}
{Smargon}, A., {Mandelbaum}, R., {Bahcall}, N., \& {Niederste-Ostholt}, M.
  2012, \mnras, 423, 856

\bibitem[{{Spergel} \& {Steinhardt}(2000)}]{Spergel00}
{Spergel}, D.~N., \& {Steinhardt}, P.~J. 2000, \it Phys. Rev. Lett.\rm, 84,
  3760

\bibitem[{Tenneti {et~al.}(2015)Tenneti, Singh, Mandelbaum, Matteo, Feng, \&
  Khandai}]{Tenneti15}
Tenneti, A., Singh, S., Mandelbaum, R., {et~al.} 2015, Monthly Notices of the
  Royal Astronomical Society, 448, 3522

\bibitem[{{West} {et~al.}(2017){West}, {de Propris}, {Bremer}, \&
  {Phillipps}}]{West17}
{West}, M.~J., {de Propris}, R., {Bremer}, M.~N., \& {Phillipps}, S. 2017,
  Nature Astronomy, 1, 0157

\bibitem[{{Wittman}(2019)}]{Wittman19analogs}
{Wittman}, D. 2019, \apj, 881, 121

\bibitem[{{Wittman} {et~al.}(2018{\natexlab{a}}){Wittman}, {Cornell}, \&
  {Nguyen}}]{Analogs2018}
{Wittman}, D., {Cornell}, B.~H., \& {Nguyen}, J. 2018{\natexlab{a}}, \apj, 862,
  160

\bibitem[{{Wittman} {et~al.}(2019){Wittman}, {Foote}, \&
  {Golovich}}]{WittmanFooteBCGalignment}
{Wittman}, D., {Foote}, D., \& {Golovich}, N. 2019, \apj, 874, 84

\bibitem[{{Wittman} {et~al.}(2018{\natexlab{b}}){Wittman}, {Golovich}, \&
  {Dawson}}]{Mismeasure2018}
{Wittman}, D., {Golovich}, N., \& {Dawson}, W.~A. 2018{\natexlab{b}}, \apj,
  869, 104

\end{thebibliography}

\end{document}